\documentclass[12pt,a4paper]{article}

\usepackage[utf8]{inputenc}
\usepackage[english]{babel}
\usepackage{graphicx}
\usepackage{float}
\usepackage{caption}
\usepackage{multicol}
\usepackage{booktabs}
\usepackage{tikz}
\usepackage[backref=page]{hyperref}
\usepackage{amsmath}
\usepackage{mathrsfs}
\usepackage{footnote}
\usepackage{tabulary}
\usepackage{subfig}
\usepackage{enumerate}
\usepackage{commath}
\usepackage{slashed}
\usepackage{rotating}
\usepackage{tikz}
\usepackage{multirow}
\usepackage{pdflscape}
\usepackage{comment}
\usepackage{authblk}
\usepackage{xcolor}
\usepackage{gensymb}

\makeatother

\begin{document}

\title{Sensitivity of accelerator-based neutrino experiments to neutrino-electron scattering radiative corrections}

\author[1]{O. G. Miranda\footnote{omr@fis.cinvestav.mx}}
\author[1]{G. Moreno-Granados\footnote{gmoreno@fis.cinvestav.mx}}
\author[2]{C. A. Moura\footnote{celio.moura@ufabc.edu.br}}
\affil[1]{Departamento de F\'{\i}sica, Centro de Investigaci\'on
  y de Estudios Avanzados del IPN, Apartado Postal 14-740 07000 Mexico,
  Distrito Federal, Mexico}
\affil[2]{Universidade Federal do ABC (UFABC), Santo Andr\'e - SP, 09210-580, Brazil}

%\date{\today}

\maketitle

\begin{abstract}
Future long-baseline experiments will measure neutrino oscillation
properties with unprecedented precision and will search for clear
signatures of {\it CP} violation in the leptonic sector. Near detectors can
measure the neutrino-electron scattering with high
statistics, giving the chance for its precise measurement. We study, in
this work, the expectations for the measurement of radiative
corrections in this process. We focus on the determination of
contributions that are exclusive to the neutrino channels,
particularly on the neutrino charge radius. We illustrate how the
perspectives in a first clear measurement of this effective quantity are encouraging.
\end{abstract}

\section{Introduction}
Since the Standard Model (SM) was proposed as an unifying electroweak
theory, the neutrino electron scattering has been proven to be a
useful test tool~\cite{Hasert:1973ff}. Its pure leptonic character has
been helpful in providing clear signatures in different predictions of the SM, such
as the existence of neutral
currents~\cite{Glashow:1961tr,Weinberg:1967tq,Glashow:1976nt,Salam:1968rm}.
Several experiments have measured the muon-neutrino scattering off
electrons, such as CHARM-II~\cite{Vilain:1994qy} and
ArgoNeuT~\cite{Anderson:2011ce}.
Currently, the neutrino electron scattering can be used to constrain
new physics, such as nonstandard
interactions~\cite{Farzan:2017xzy,Ohlsson:2012kf,Miranda:2015dra} and
a neutrino magnetic
moment~\cite{Giunti:2014ixa,Canas:2015yoa,Daraktchieva:2005kn}.

Regarding precision tests of the SM, measurements other than neutrino
electron scattering have proven to be a powerful tool.  For example,
precise measurements of the weak mixing angle are made at high
energies in $e^{+} e^{-}$ and $p^{+}p^{-}$ collisions that can be
extrapolated to lower energies.
However, a better determination of the SM parameters at low-energy
experiments can give direct proof of the model in this energy
region.
A current subject of interest is the precise determination of the weak
mixing angle at a low momentum transfer that is performed, for example,
with atomic parity violation
experiments~\cite{Marciano:1990dp,Kumar:2013yoa}. Another scenario where we can also
test this energy regime is the coherent elastic neutrino nucleus
scattering~\cite{Tomalak:2020zfh,Cadeddu:2018izq,Canas:2018rng}.

Current measurements of the weak mixing angle through neutrino
electron scattering still have large uncertainties due to the small
cross section and the difficulty of generating enough statistics from
relatively small neutrino fluxes~\cite{Vilain:1994qy,Deniz:2009mu}.
In this case, the uncertainties do not allow us to distinguish if the
radiative corrections predicted by the theory are affecting
measurements of the weak mixing angle beyond the errors.  However, the
search for the existence of a {\it CP}-violating phase in the neutrino
sector has motivated the construction of long-baseline experiments
that predict intense neutrino beams. That opens the possibility to
measure neutrino electron scattering in the facility near detector
(ND), provided the systematic uncertainties can be kept under
control. The impact of radiative corrections in this context has been
studied, for example, in
Refs.\cite{deGouvea:2019wav,Marshall:2019vdy}.

A confirmation of the predicted value of the weak mixing angle at low
energies is an important test that these experiments can perform.
Moreover, radiative corrections to neutrino-electron
  scattering have flavor-dependent contributions that are particular to
  neutrino interactions. Usually, this flavor-dependent correction is
  referred to as the neutrino charge radius and represents, by itself, a new
  test of the SM that the new generation of long-baseline neutrino
   experiments could provide.
The neutrino charge radius
leads to a shift in the effective value of the weak mixing angle,
making them entangled.
It has been studied
thoroughly~\cite{Sarantakos:1982bp,Lucio:1983mg,CabralRosetti:1999ad,Fujikawa:2003ww,Papavassiliou:2003rx},
and there has been a long discussion about its correct definition.
Recently, this discussion has lead to a
definition~\cite{CabralRosetti:1999ad,Papavassiliou:2003rx} of an
effective neutrino charge radius that is gauge independent and
fulfills all the necessary physical
properties~\cite{Papavassiliou:2003rx}. The discussion on this topic
makes it even more interesting the possibility that the neutrino
charge radius would contribute to an observable displacement of the
effective value of the weak mixing angle.

In this work we focus in the possibility that the ND at long baseline
neutrino experiments may help to test such a quest. For definiteness,
we center our discussion on the DUNE proposal, considering a
PRISM-like detector~\cite{Hongyue:2018jae,DeRomeri:2019kic}. It may also
be interesting to study this phenomenology in other
configurations. The short-baseline neutrino program at Fermilab might
as well be another configuration to study these effects. In
particular, SBND~\cite{Brailsford:2017rxe,Mcconkey:2018gpr} and
ICARUS~\cite{Antonello:2015lea,Farnese:2019xgw} are expected to take
data in the near future, and a dedicated program to measure neutrino
electron scattering may be of interest.

\section{Radiative corrections and neutrino charge radius}

In this section, we introduce the main characteristics of the
muon-(anti)neutrino electron scattering, $\nu_\mu(\bar{\nu}_\mu)e^-$,
both at tree level and with the addition of radiative
corrections. This is a neutral current $Z$-mediated process that is
clean in the sense that it involves only leptons; therefore, quantum
chromodynamics related physics is absent. The process allows, at least
in principle, a precise test of the SM at low energies, in particular,
the consistency of the weak mixing angle and the possible existence of
a neutrino charge radius.

The differential cross section for the $\nu_\mu e^-$ scattering at
tree level is given by
\begin{align}
  \label{eq:cs:tree}  
  \frac{d\sigma}{dT}& =\frac{2m_eG^2_F}{\pi}\left\lbrace g_L^2  + g_R^2\left(1-\frac{T}{E_{\nu}} \right)^2 -g_R g_L m_e \frac{T}{E_{\nu}^2} \right\rbrace ,     
\end{align}
where $m_e$ is the electron mass, $G_F$ is the Fermi constant, $T$ is
the electron kinetic energy of recoil, and $E_{\nu}$ is the incoming
neutrino energy. The coupling constants $g_L$ and $g_R$ are defined 
at tree level, as
\begin{subequations}
  \label{eq:couplings}
  \begin{equation}
    \label{eq:gLtree}
    g_L= \frac{1}{2} - \sin^2\theta_W
  \end{equation}
  and
  \begin{equation}
    \label{eq:gRtree}
    g_R= - \sin^2\theta_W \,,
  \end{equation}
  where $\theta_W$ is the weak mixing angle.
\end{subequations}

\subsection{Radiative corrections}
\label{subsec:rc}
\begin{figure}[htb]
  \subfloat[QED radiative correction.\label{fig:QEDcorrec}]{%
    \includegraphics[width=0.45\textwidth]{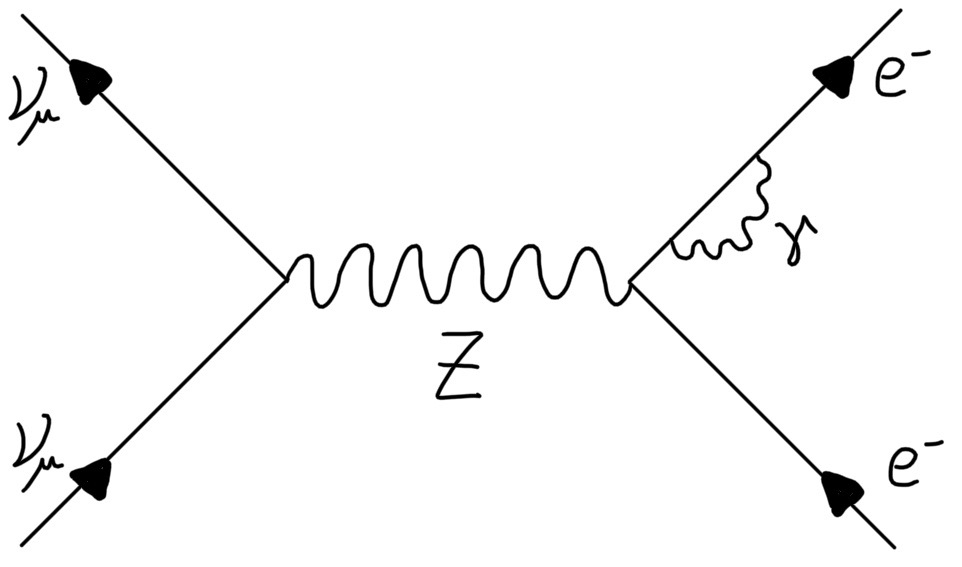}}
  \hfill
  \subfloat[EW radiative correction.\label{fig:EWcorrec}]{% 
    \includegraphics[width=0.45\textwidth]{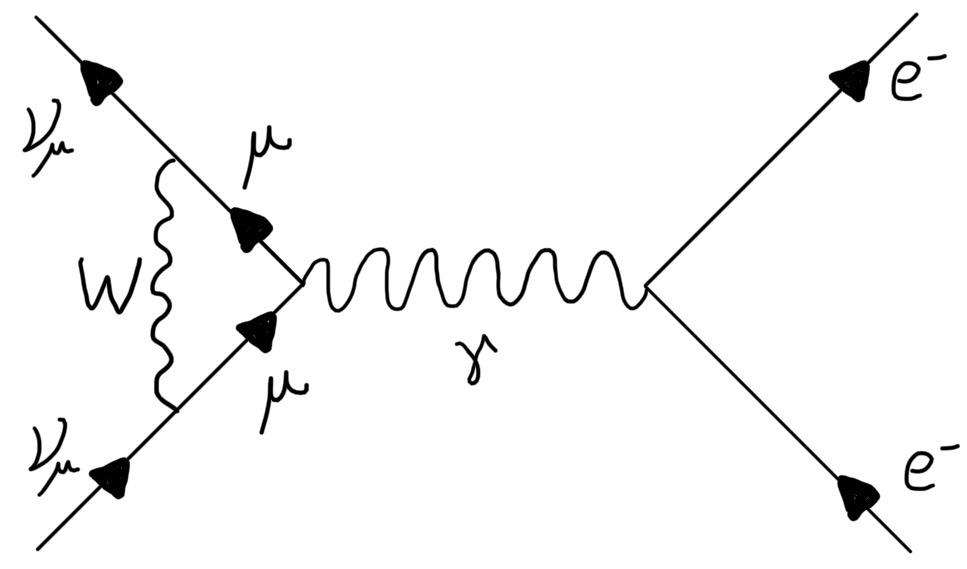}}
  \caption[]{Feynman diagrams representing high order radiative
    corrections from (a) QED, $e\gamma e$ vertexes and (b) EW, $\mu
    W\nu_\mu$ vertexes.  }
  \label{fig:correcciones}
\end{figure}

Radiative corrections in $\nu_\mu e$ scattering have been extensively
studied~\cite{Bahcall:1995mm,Passera:2000ug,Passera:2001gh,Sirlin:2012mh,Ferroglia:2003wa,Erler:2017knj,Zyla:2020zbs,Tomalak:2019ibg,Bischer:2018zcz}. They
can be divided into two different groups, depending on their dynamic
origin: (a) quantum electrodynamic (QED) corrections, that involve,
e.g., the creation and absorption of photons in the electronic
current, as illustrated in Fig.~\ref{fig:QEDcorrec}. (b) Electroweak
(EW) corrections, due to the exchange of $W$ and $Z$ bosons, for
instance, the one shown in Fig.~\ref{fig:EWcorrec}. The EW
corrections, as we discuss later, include the neutrino charge radius.

The expression considering QED and EW radiative corrections for the
$\nu_\mu e^-$ differential cross section takes the form,
\begin{align}
  \label{eq:cs:ewqed}  
  \frac{d\sigma^\prime}{dT}& =\frac{2m_eG^2_F}{\pi}\left\lbrace g_L^{\prime 2}(T)\left[1+ \frac{\alpha}{\pi}f_{-}(z)\right]  + g_R^{\prime 2}(T)\left(1-\frac{T}{E_{\nu}} \right)^2\left[1+ \frac{\alpha}{\pi}f_{+}(z)\right] \right. \nonumber\\
  &\qquad {} \left. -g^\prime_R(T)g^\prime_L(T) m_e \frac{T}{E_{\nu}^2}\left[1+ \frac{\alpha}{\pi}f_{+-}(z)\right]\right\rbrace ,
\end{align}
where the functions $f_{+}(z)$, $f_{-}(z)$, and $f_{+-}(z)$ account
for the QED corrections that depend on $z= T/E_{\nu}$. $\alpha$ is the
fine-structure constant. The expressions for these
functions~\cite{Bahcall:1995mm} are given in Appendix~\ref{appen.A}.  The
values of $f_{+}(z)$, $f_{-}(z)$, and $f_{+-}(z)$ present important
variations with the neutrino energy in the range under consideration.
For the antineutrino cross section, the $g^\prime_{L,R}$ couplings
must be interchanged like $g^\prime_L \leftrightarrow g^\prime_R$,
while the three functions, $f(z)$, are preserved.

The coupling constants now include the EW corrections in the following way:
\begin{subequations}
  \label{eq:ew:gs:mu}
  \begin{equation}
    \label{eq:ew:gL:mu}
    g^\prime_L(T)=\rho_{\rm NC}\left[ \frac{1}{2} - \kappa_{\nu_l}(T)\sin^2\theta_W^{(m_Z)}\right]
  \end{equation}
  and
  \begin{equation}
    \label{eq:ew:gR:mu}
    g^\prime_R(T)= - \rho_{\rm NC}\kappa_{\nu_l}(T)\sin^2\theta_W^{(m_Z)} \,,
  \end{equation}
\end{subequations}
  where $\rho_{\rm NC}$ and $\kappa_{\nu_l}(T)$ are defined in
  Eqs. (\ref{rho}) and (\ref{kappa:sirlin}) below.  $m_Z$ is the
  $Z$ boson mass and $\sin^2\theta_W^{(m_Z)}$ is $\sin^2\theta_W$
  calculated at the $m_Z$ scale. We follow closely the
  analytic expressions reported in
  Ref.~\cite{Sirlin:2012mh}.  This approach, as we discuss later,
  allows us to confirm that, as expected, the EW corrections do not
  have important variations in the energy range of our interest.
  
\begin{equation}
  \label{rho}
  \rho_{\rm NC}= 1 + \frac{\hat{\alpha}}{4\pi \hat{s}^2} \left\lbrace \frac{3}{4\hat{s}^2}\ln c^2 - \frac{7}{4} + \frac{2\hat{c}_Z}{\hat{c}^2} + \frac{3}{4}\xi \left[ \frac{\ln\left( \frac{c^2}{\xi} \right)}{c^2 - \xi} + \frac{1}{c^2}\frac{\ln \xi}{1-\xi} \right] + \frac{3}{4}\frac{m_t^2}{m_W^2} \right\rbrace ,
\end{equation}
where $s$ and $c$ stand for sine and cosine of $\theta_W$,
respectively. Hat over the parameters indicate their values calculated
at the $m_Z$
scale. $\hat{c}_Z=\frac{19}{8}-\frac{7}{2}\hat{s}^2+3\hat{s}^4\,$,
$\xi=\frac{m_H^2}{m_Z^2}\,$, and $m_{H,t,W}$ are the masses of the
Higgs boson, the top quark, and the $W$ boson respectively.  Rho has
the numerical value $\rho_{\rm NC} = 1.014032$.

\begin{align}
  \label{kappa:sirlin}
  \kappa_{\nu_l}(q^2) 
  &= 1- \frac{\alpha}{2\pi\hat{s}^2}\left[ \sum_i \left( C_{3i}Q_i-4\hat{s}^2Q_i^2\right)J_i(q^2) - 2J_l(q^2) \right. \nonumber\\
    &\qquad {} \left.+ \ln c \left( \frac{1}{2} -7\hat{c}^2 \right) + \frac{\hat{c}^2}{3} + \frac{1}{2} + \frac{\hat{c}_{\gamma}}{\hat{c}^2}\right] ,
\end{align}
where
$C_{3i}$ is twice the third component of weak isospin,
$Q_i$ represents the electric charge,
$\hat{c}_\gamma=\frac{19}{8}-\frac{17}{4}\hat{s}^2+3\hat{s}^4\,$,
$q^2=-2m_eT$ is the squared four-momentum transfer, and
  \begin{equation}
    \label{kap:sirlin:Ji}
    J_i(q^2)= \int_0^1 x(1-x)\ln\left( \frac{m_i^2-q^2x(1-x)}{m_Z^2}\right) dx\,,
  \end{equation}
  where $m_i$ is the mass of the $i$th fermion.  The sum in
  Eq.~\eqref{kappa:sirlin} includes all the charged
  fermions, and we consider an additional factor of 3 for
  quarks (due to the color degree of freedom).

The flavor dependence of the incident neutrino is contained in the
$2J_l(q^2)$ term. For a $\nu_\mu$ flux, we have $2J_\mu(q^2)$.  We can
have a first general idea of the different dependence on EW
corrections for neutrino and antineutrino electron scattering by
considering the simple case of a monoenergetic neutrino beam and
focus on the effect of $\kappa_{\nu_\mu}$.  Now, the cross section is
given by an equation similar to Eq.~(\ref{eq:cs:tree}), but corrected
with the coupling constants,

\begin{subequations}
  \label{eq:ew:gs:ex}
\begin{equation}
    \tilde{g}_L\approx \frac{1}{2} - \kappa_{\nu_\mu} x 
    %\hspace{1cm}
\end{equation}
    and
\begin{equation}    
    %\hspace{1cm}  
    \tilde{g}_R\approx - \kappa_{\nu_\mu} x ,
\end{equation}
\end{subequations}
with $x=\sin^2\theta_W$ for short.

The differences between the aforementioned differential cross section,
considering EW radiative corrections (${d\sigma^\prime_{\rm EW}}/{dT}$) and
the differential cross section at tree level, Eq.~(\ref{eq:cs:tree}),
for neutrino and antineutrino, are, respectively,
\begin{subequations}
\begin{equation}
  \Delta\sigma_{\nu_\mu e} \equiv \frac{d\sigma^\prime_{\rm EW}}{dT} - \frac{d\sigma}{dT} \sim \Delta g_L + \Delta g_R \left( 1-2\frac{T}{E_\nu} + \frac{T^2}{E_{\nu}^2} \right) - \Delta g_{R,L}m_e\frac{T}{E_{\nu}^2} 
\end{equation}
and
\begin{equation}
  \Delta\sigma_{\bar{\nu}_\mu e} \equiv \frac{d\bar{\sigma}^\prime_{\rm EW}}{dT} - \frac{d\bar{\sigma}}{dT} \sim \Delta g_R + \Delta g_L \left( 1-2\frac{T}{E_\nu} + \frac{T^2}{E_{\nu}^2} \right) - \Delta g_{R,L}m_e\frac{T}{E_{\nu}^2} \,,
\end{equation}
\end{subequations}
where $\Delta g_L$, $\Delta g_R$, and $\Delta g_{R,L}$ represent the following differences:
\begin{subequations}
\begin{align}
  \label{Delta_gL}
  \Delta g_L &\equiv \tilde{g}_L^2 - g_L^2 = (\kappa^2_{\nu_\mu} - 1) x^2 \left[1 - \frac{1}{(\kappa_{\nu_\mu}+1)x}\right] \,,
\end{align}
\begin{align}
  \label{Delta_gR}
  \Delta g_R &\equiv \tilde{g}_R^2 - g_R^2 = (\kappa^2_{\nu_\mu} -1) x^2 \,,
  \end{align}
  and
\begin{align}
  \label{Delta_gRL}
  \Delta g_{R,L} &\equiv \tilde{g}_R\tilde{g}_L - g_Rg_L =
  (\kappa^2_{\nu_\mu} - 1) x^2 \left[1 - \frac{1/2}{(\kappa_{\nu_\mu}+1)x}\right] \,. 
\end{align}
\end{subequations}
Given $\kappa_{\nu_\mu}\approx 1$ and $x\approx 1/4$, we have $(\kappa_{\nu_\mu} +1)x\approx 1/2$, which implies
\begin{align}
\Delta g_L\approx - \Delta g_R \qquad {\rm and} \qquad \Delta g_{L,R} \approx 0\,,
\end{align}
and, therefore:
\begin{equation}
  \label{Delta:sigma:nue}
  \Delta\sigma_{\bar{\nu}_\mu e} \approx - \Delta\sigma_{\nu_\mu e} \approx
    \Delta g_R \left(-2 + \frac{T}{E_{\nu}} \right)\frac{T}{E_\nu}\,.
\end{equation}

\begin{figure}[htb]
\centering
     \includegraphics[width=1.0\textwidth]{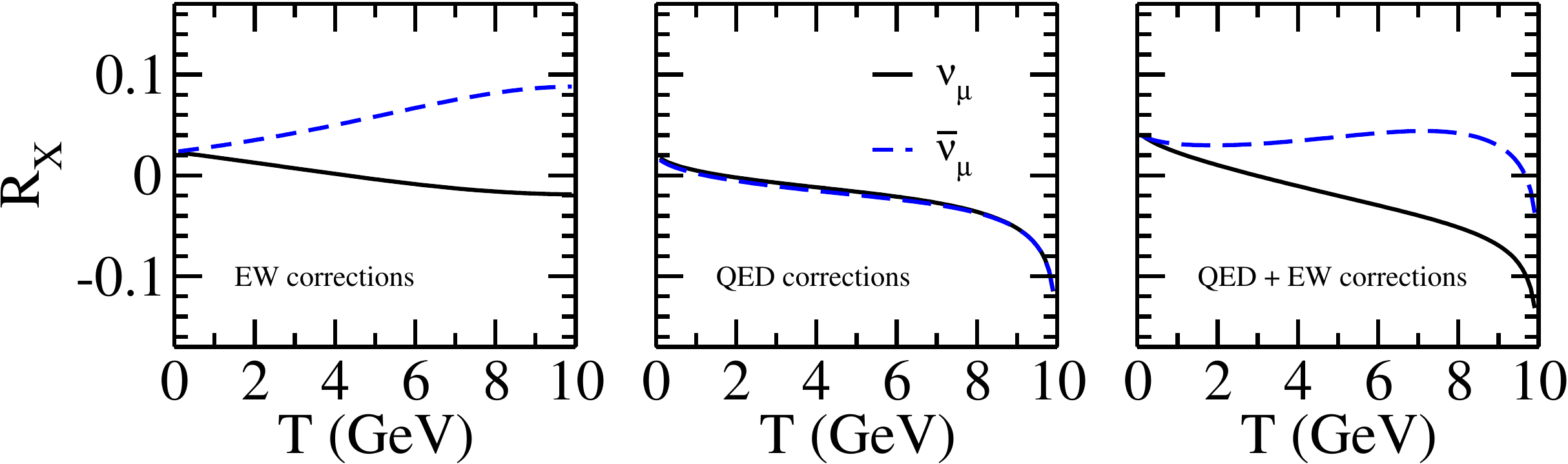}
  \caption[]{\small Comparison of the ratio of radiative corrections
    for neutrino and antineutrino beam modes, for a fixed neutrino
    energy of 10~GeV. a) Contribution of EW corrections, b)
    contribution of QED corrections, and c) total contributions.}
	\label{fig:corrections:nuAntinu}
\end{figure}

This result shows that there is an asymmetric relation between the EW
radiative corrections for neutrino and antineutrino scattering.
Conversely, if we consider the QED corrections effect, the
relative deviation from the tree level is the same for neutrino and
antineutrino.  These behaviors are illustrated in
Fig.~\ref{fig:corrections:nuAntinu}, where the relative contributions,
Eq.~\eqref{eq:ratio}, of the different groups of corrections are
displayed for a hypothetical monoenergetic neutrino beam of 10~GeV.
The deviation from the tree level differential cross section is
defined as the ratio,
\begin{equation}
  \label{eq:ratio}
  R_{\rm X}:=\dfrac{\frac{d\sigma^\prime_{\rm X}}{dT}-\frac{d\sigma}{dT}}{\frac{d\sigma}{dT}} \,,
\end{equation}
where X denotes the inclusion of either EW, QED, or both corrections
at the same time. Although the behavior of this ratio was calculated
for a fixed neutrino energy, the qualitative behavior persists for the
neutrino beam spectrum.

The two different effects (EW + QED corrections, Fig.~\ref{fig:corrections:nuAntinu}c) change the
antineutrino electron scattering cross section, resulting practically always in an
increment.
This is in opposition to the neutrino case, in which the radiative
corrections increase the cross section only in the low energy range,
below $\approx 2$~GeV for our study case depicted in
Fig.~\ref{fig:corrections:nuAntinu}, while in the higher energy range
the cross section decreases. This behavior will be relevant when
studying specific experimental setups, as we evince in
Sec.~\ref{sec:dune}.

\subsection{Neutrino charge radius}
We can now take a careful look at the contribution of
$\kappa_{\nu_l}(q^2)$, defined in Eq.~\eqref{kappa:sirlin}.
Depending on the particular process, e.g., for different neutrino flavors, the correction has different values. 
We can decompose this expression into two parts. The first one,
$\kappa_{\nu}(q^2)$, is a common contribution for all the neutrino
flavors:

\begin{align}
  \label{kappa:sirlin:sinJi}
  \kappa_{\nu}(q^2) =
  1 - \frac{\alpha}{2\pi\hat{s}^2}\left[ \sum_i \left( C_{3i}Q_i-4\hat{s}^2Q_i^2\right)J_i(q^2) + \ln c \left( \frac{1}{2} -7\hat{c}^2 \right) + \frac{\hat{c}^2}{3} +  \frac{\hat{c}_{\gamma}}{\hat{c}^2}\right] . 
\end{align}	
In the energy region of interest for this work, this
  contribution takes the value $\kappa_{\nu}(q^2) = 1.017$.

The second contribution is flavor dependent,
\begin{align}
  \label{kappa:sirlin:Jnu}
  - \frac{\alpha}{2\pi\hat{s}^2}\left[ - 2J_l(q^2) + \frac{1}{2} \right] , 
\end{align}
and its numerical value in this region is $-0.025$.

\begin{figure}[htb]
\centering
    \includegraphics[width=0.7\textwidth]{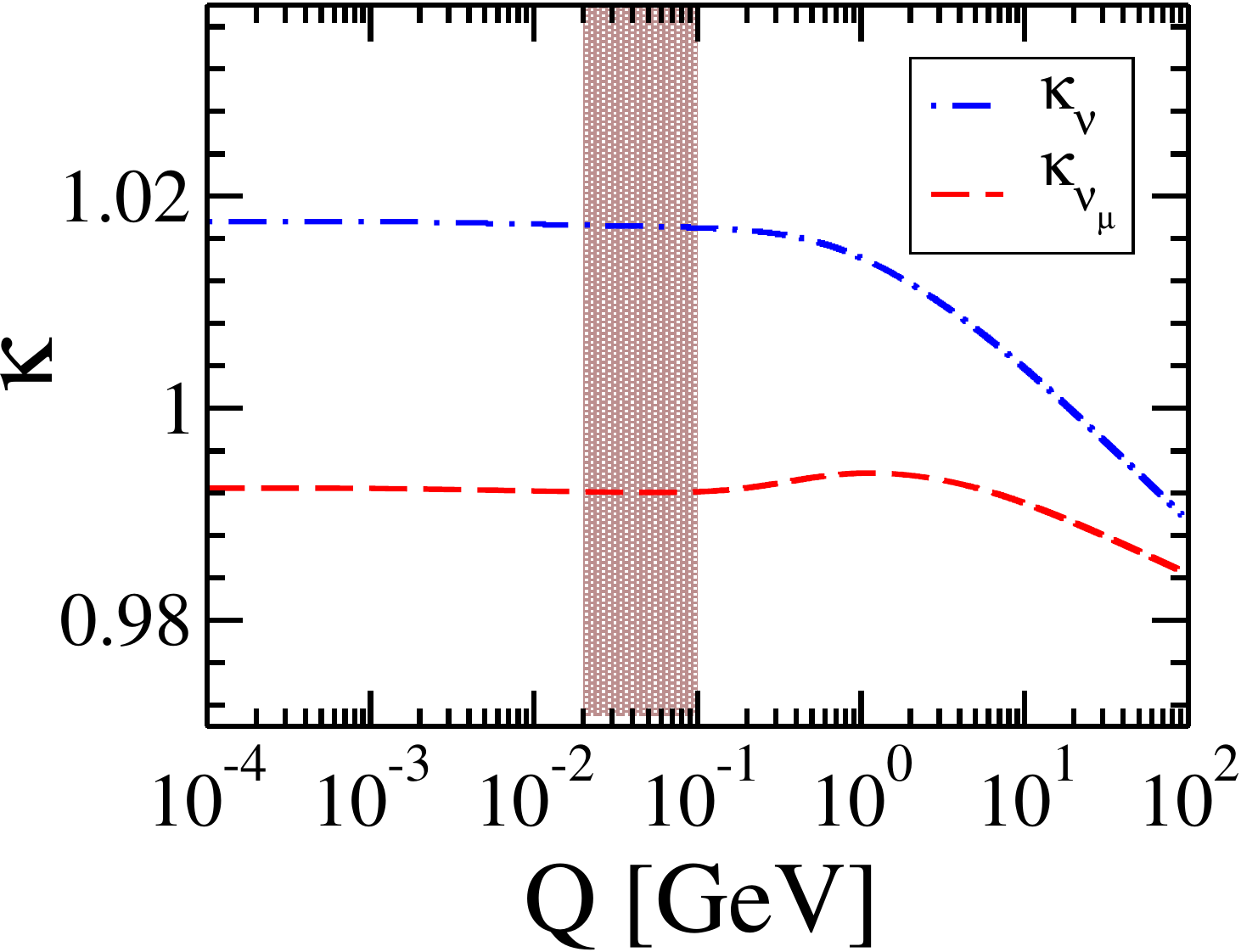}
    \caption[]{\small $\kappa_{\nu}$ and $\kappa_{\nu_\mu}$ as
      functions of $Q$.  The dot-dashed blue line represents
      $\kappa_\nu$, Eq.~\eqref{kappa:sirlin:sinJi}, and the dashed red
      line represents $\kappa_{\nu_\mu}$,
      Eq.~\eqref{kappa:sirlin}. The shaded area represents the
      electron recoil energy (T) where we investigate the effect of
      radiative corrections and the experimental sensitivity to the
      neutrino charge radius.}
\label{fig:kappas}
\end{figure}

This is responsible for the difference of around $3\%$ between
$\kappa_{\nu_\mu}$, Eq.~\eqref{kappa:sirlin}, and $\kappa_\nu$, Eq.~\eqref{kappa:sirlin:sinJi}, shown in
Fig.~\ref{fig:kappas}. When $Q\equiv \sqrt{-q^2}$ tends to zero, the
values of $\kappa$, i.e., the value of the flavor dependent part,
Eq.~\eqref{kappa:sirlin:Jnu}, remains constant. This encompasses the
energy range of our interest in this work, limited by the shaded
vertical band in the figure.

We turn our attention to
$- 2J_l(q^2) + \frac{1}{2}$, from Eq.~\eqref{kappa:sirlin:Jnu}, which when evaluated in $q=0$ becomes
	\begin{equation}
	\label{eq:part2}
	- 2J_l(0) + \frac{1}{2}=\frac{1}{6}\left[ 3-2\ln\left( \frac{m^2_l}{m^2_Z} \right) \right] .
	\end{equation}
This quantity~\footnote{The right-hand side of Eq.~\eqref{eq:part2}
  can be written in terms of $m_W$ adding $\frac{1}{3}\ln\left(
  \frac{m^2_W}{m^2_Z} \right)$ to it.  In the shaded region of
  Fig.~\ref{fig:kappas}, we have $\kappa_{\nu_\mu}^{(m_Z)} = 0.9921 $
  and $\kappa_{\nu_\mu}^{(m_W)} = 0.9925$, only a $\sim 0.04$\%
  difference.} is usually associated with the neutrino charge radius
(NCR),
  \begin{equation}
    \label{nu:char:rad}
    \left\langle r^2_{\nu_l} \right\rangle = \frac{G_F}{4\sqrt{2}\pi^2} \left[ 3-2\ln\left( \frac{m^2_l}{m^2_W} \right) \right],
  \end{equation}
for which the reported value for the $\mu$ flavor is $\left\langle
r^2_{\nu_\mu}
\right\rangle=2.4\times10^{-33}$~cm$^2$~\cite{Giunti:2014ixa}.

We can also separate the couplings $g^\prime_L(T)$ and $g^\prime_R(T)$
into two parts, one independent of the incoming neutrino flavor and the
other in terms of the NCR as~\cite{Giunti:2014ixa},
  \begin{equation}
    \label{eq:gL:gR:w-woCR}
    g_{L,R}^{\prime(\nu_{\mu},e)}(T) \sim g_{L,R}^{\prime(\nu,e)}(T)+\left[ \frac{2}{3}m_W^2 \left\langle r^2_{\nu_\mu} \right\rangle \right]\sin^2\theta_W^{(m_Z)} ,
  \end{equation}
where the numerical value of $\left[ \frac{2}{3}m_W^2
  \left\langle r^2_{\nu_\mu} \right\rangle
  \right]\sin^2\theta_W^{(m_Z)}$ is $\approx 0.0058$.

The numerical values of $\kappa$ and the couplings are reported in
Table~\ref{tab:kappa:gL:gR}. The first row shows the flavor
independent values, i.e., without the term containing the NCR. The
second row shows the values including the NCR term.
\begin{table}[htb]
  \centering
  \caption[]{\small Numerical value of $\kappa$ (evaluated at $q^2=0$) and of the couplings $g^\prime_L(T)$ and $g^\prime_R(T)$, depending on the inclusion of the neutrino charge radius term.}
    \begin{tabulary}{1.0\textwidth}%{ | p{5cm} | p{5cm} | p{5cm} |}
    { | c | c | c | c |}\hline
    NCR & $\kappa$ & $g^\prime_L$ & $g^\prime_R$ \\ \hline
no  & 1.0176 & 0.2684 & -0.2386 \\ \hline
yes & 0.9925 & 0.2743 & -0.2327 \\ \hline
    \end{tabulary}
  \label{tab:kappa:gL:gR}
\end{table}

\section{The DUNE case}
\label{sec:dune}

The Deep Underground Neutrino Experiment (DUNE)~\cite{Abi:2020wmh} is
part of one of the most ambitious neutrino experimental programs in
the world, consisting of two detectors separated by a baseline of
approximately 1300 km. There will be a 40 kt far detector
(FD)~\cite{Abi:2020evt} in South Dakota and a near detector
(ND)~\cite{Hongyue:2018jae} in Illinois at the Fermi National
Accelerator Laboratory. The ND measures the flux spectrum with no
oscillation for all the neutrino types coming from the beam. To
achieve DUNE measurement goals of precision, the ND must provide
constraints on the systematic uncertainties such as the absolute and
relative flux, nuclear effects, and neutrino type determination. It is expected to achieve very precise measurements of neutrino
interactions. There are a few ND design options, but in this work, we assume the LArTPC design, which uses the same technology as the FD.

In order to minimize the systematic uncertainties in the flux, cross
section, and detector effects in the energy spectrum, a movable near
detector concept was proposed, called
DUNE-PRISM~\cite{vilela:2018,DeRomeri:2019kic}. With PRISM, it is
possible to collect data at several off axis angles up to a maximum of
$3.6^\circ$, exposing the ND to different fluxes and spectra.

To explore DUNE-PRISM sensitivity to the NCR, we calculate the expected
number of events generated from $\nu_\mu e^-$ scattering for on axis
and different off axis beam angles.  Then we analyze what can be the
best angular window and energy range to measure differences in the
number of events related to the radiative corrections, especially the
NCR effect.

\begin{figure}[htb]
\centering
   \subfloat[$\nu_\mu$ beam mode \label{fig:nu:flux}]{%
    \includegraphics[width=0.5\textwidth]{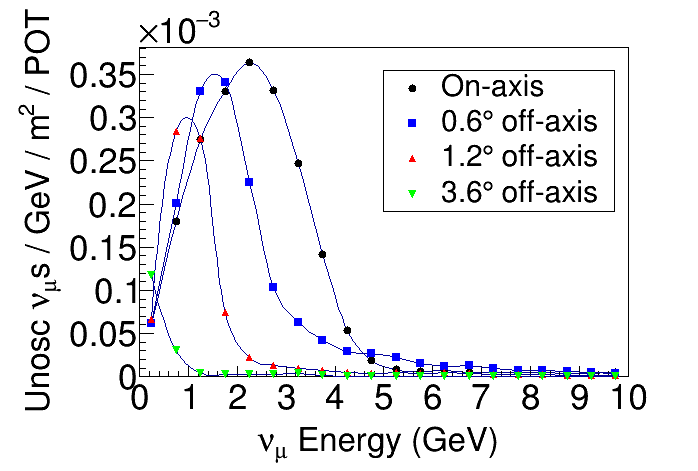}}
  \hfill
   \subfloat[$\bar{\nu}_\mu$ beam mode \label{fig:antinu:flux}]{% 
     \includegraphics[width=0.5\textwidth]{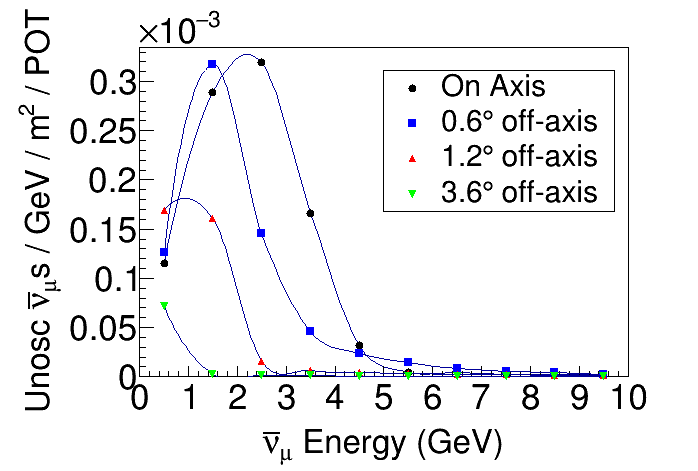}}
  \caption[]{\small Fluxes at several off axis
    locations~\cite{DUNEFluxes:2017}. Neutrino mode (a), on the left
    side, and antineutrino mode (b), on the right side. The symbols
    represent the simulated data, and the lines show their
    interpolation.  }
	\label{fig:fluxes:NumuAntiNumu}
\end{figure}

For this purpose, we consider the predicted neutrino energy spectra
for the different incident beam angles.  We consider the fluxes
reported in Ref.~\cite{DUNEFluxes:2017}.  These fluxes are shown in
Fig.~\ref{fig:fluxes:NumuAntiNumu} with the simulated data represented
by different symbols for each different beam direction and with the
data interpolation lines.

The number of targets in the detector corresponds to the number of electrons in the total mass of liquid argon, considered to be 75 t~\cite{DeRomeri:2019kic}. The
experiment is expected to run for 3.5 years in the neutrino mode and
the same period in the antineutrino mode.  Given these assumptions, we
compute the expected number of events without or with radiative
corrections, in which case, we can separate and distinguish the NCR
contribution to the corrections.

For each PRISM axis configuration, in terms of
  angular location, we must compute the average cross section. That is
  given by the integral from the threshold electron recoil energy,
  $T_{min}$, up to a maximum kinetically allowed value, $ T_{max}
  \approx E_{\nu}^{max} $, so
\begin{equation}
  \label{eq:cs:TOTAL}
  \sigma = \int_{T_{min}}^{T_{max}} f(T) dT,
\end{equation}
where $f(T)$ is the integral of the differential cross
section, $ \frac{d \sigma}{dT} \left(T, E_{\nu} \right) $, times the
corresponding neutrino flux, $ \lambda (E_{\nu }) $,
\begin{equation}
  \label{eq:integral:cs:flux}
  f(T)=\int_{E_{\nu}^{min}(T)} ^{E_{\nu}^{max}} \frac{d\sigma}{dT}\left(T,E_{\nu} \right) \lambda(E_{\nu}) dE_{\nu}\,,
\end{equation}
where $E_\nu^{min}(T)$ is the minimum neutrino energy considered and given by the detector's electron energy threshold.

Once we compute the cross section, Eq.~\eqref{eq:cs:TOTAL}, it is
necessary to take into account the detector exposure to obtain the
number of events,
\begin{equation}
\label{eq:eventnumber}
N = \sigma \times \mathscr{C} \,,
\end{equation}
where $\mathscr{C}$ is the exposure.  It takes into account the number
of target electrons in the detector, the number of protons on target
per year of $1.1\times10^{21}$~POT/year~\cite{DUNEFluxes:2017}, and
3.5 years in the neutrino beam mode plus 3.5 years in the antineutrino
beam mode.

As discussed in Sec.~\ref{subsec:rc}, it is expected that, when
radiative corrections are considered, the antineutrino electron
scattering cross section have an increase with respect to the
tree level calculation. This results in an increase in the expected
number of events as well.  It is not the case for the neutrino mode in
which, as we discuss in the next section, a decrease in the expected
number of events is predicted for electron recoil energies above
approximately $0.7$~GeV.  

\subsection{Results and discussion}
\label{ssec:res}

\begin{table}[htb]
  \centering
  \caption[]{\small The total number of events from $\bar{\nu}_\mu e$
    scattering for an energy range of 0.2 to 10 GeV, considering the
    tree level and radiative corrections with and without neutrino
    charge radius (NCR). The first column shows the DUNE-PRISM axis
    location. $\sigma_{stat}$ is the statistical error and $\Delta$ is
    the difference between the number of events calculated at tree level and with radiative corrections. See text for details.}
    \begin{tabulary}{1.0\textwidth}%{ | p{5cm} | p{5cm} | p{5cm} |}
    { | c | c | c | c | c | c | c |}\hline
    & \multicolumn{6}{ |c| }{Number of $\bar{\nu}_\mu$ Events} \\ \hline
    & & & \multicolumn{2}{ |c| }{Without NCR} & \multicolumn{2}{ |c| }{With NCR} \\ \hline
    Axis location & Tree level & $\sigma_{stat}$ & EW+QED & $\Delta$ & EW+QED & $\Delta$ \\ \hline
0.0$\degree$ & 18775 & 137 & 19931  & 1156 & 19447  & 672 \\ \hline
0.6$\degree$ & 11969 & 109 & 12715  & 746 & 12402   & 433 \\ \hline
1.2$\degree$ & 3993  & 63 & 4251    & 258 & 4141    & 148 \\ \hline
1.8$\degree$ & 1181  & 34 & 1260    & 79 & 1226     & 45 \\ \hline
2.4$\degree$ & 645   & 25 & 689     & 44 & 670      & 25 \\ \hline
3.0$\degree$ & 437   & 21 & 467     & 30 & 454      & 17 \\ \hline
3.6$\degree$ & 315   & 18 & 336     & 21 & 327      & 12 \\ \hline				
    \end{tabulary}
  \label{tab:totalEvents:AntiNumu:0.2}
\end{table}

We show our results considering two different electron recoil energy
thresholds. The first is 0.2~GeV, and the second is 0.7~GeV. The second
threshold is used to maximize the difference in the number of events
between the tree level calculation and the radiative corrections for
the neutrino beam mode. For on axis neutrino beam, the effect of
radiative corrections is opposite below and above the threshold of
$\sim 0.7$~GeV. When the differential cross section is integrated, this
ends up diminishing the total event number difference. This is why
considering this crossing point as the threshold helps to enlarge the
difference that we need to detect in the case of neutrino
scattering.
Considering this crossing point as the threshold helps to enlarge the
  difference between the tree level and the one loop level predictions.
For antineutrinos, radiative corrections increase the
expected number of events, independent of the energy range observed.

\begin{table}[htb]
  \centering
  \caption[]{\small The total number of events from $\nu_\mu e$
    scattering for an energy range of 0.2 to 10 GeV, considering the
    tree level and radiative corrections with and without neutrino
    charge radius. The first column shows the DUNE-PRISM axis
    location. $\sigma_{stat}$ is the statistical error and $\Delta$ is
    the difference between the number of events calculated at tree level and with radiative corrections. See text for details.}
    \begin{tabulary}{1.0\textwidth}%{ | p{5cm} | p{5cm} | p{5cm} |}
    { | c | c | c | c | c | c | c |}\hline
    & \multicolumn{6}{ |c| }{Number of $\nu_\mu$ Events} \\ \hline
    & & & \multicolumn{2}{ |c| }{Without NCR} & \multicolumn{2}{ |c| }{With NCR} \\ \hline
    Axis location & Tree level & $\sigma_{stat}$ & EW+QED & $\Delta$ & EW+QED & $\Delta$ \\ \hline
0.0$\degree$ & 27134 & 165 & 25859 & -1275 & 26567 & -567 \\ \hline
0.6$\degree$ & 18099 & 135 & 17243 & -856  & 17712 & -387 \\ \hline
1.2$\degree$ & 5884  & 77  & 5589  & -295  & 5749  & -135 \\ \hline
1.8$\degree$ & 2600  & 51  & 2466  & -134  & 2538  & -62 \\ \hline
2.4$\degree$ & 1397  & 37  & 1324  & -73   & 1364  & -33 \\ \hline
3.0$\degree$ & 711   & 27  & 674   & -37   & 694   & -17 \\ \hline
3.6$\degree$ & 440   & 21  & 418   & -22   & 430   & -10 \\ \hline				
    \end{tabulary}
  \label{tab:totalEvents:Numu:0.2}
\end{table}

In Table~\ref{tab:totalEvents:AntiNumu:0.2} and
Table~\ref{tab:totalEvents:Numu:0.2}, we summarize the results for
antineutrino and neutrino events, respectively. The energy range
considered in the calculation is 0.2~GeV to 10.0~GeV. We give the
results for the DUNE-PRISM axis location from 0$\degree$ to 3.6$\degree$ in
intervals of 0.6$\degree$ and show the NCR's contribution in the
radiative corrections.
Considering the NCR, the difference in the number of events, $|\Delta|$, in comparison with the tree level calculation, is larger for antineutrino than for neutrino.  More
importantly, we see in these tables that
the difference in the number of events is bigger than the statistical error for off axis angles equal to or smaller than $1.8\degree$. In particular, for the on axis case, the statistical error is remarkably
small in comparison to the difference in the number of
events. Therefore, if the systematic uncertainties can be under
control, a determination of the NCR may be possible.

\begin{table}[htb]
  \centering
  \caption[]{\small The total number of events for $\bar{\nu}_\mu$ and
    $\nu_\mu$ beam modes, on axis, within the energy range from 0.7 to
    10 GeV, considering the tree level and radiative corrections with
    and without neutrino charge radius. $\sigma_{stat}$ is the
    statistical error and $\Delta$ is the difference between the
    number of events calculated at tree level and with radiative
    corrections. See text for details.  }
    \begin{tabulary}{1.0\textwidth}%{ | p{5cm} | p{5cm} | p{5cm} |}
    { | c | c | c | c | c | c | c |}\hline
    & \multicolumn{6}{ |c| }{Number of Events} \\ \hline
    & & & \multicolumn{2}{ |c| }{Without NCR} & \multicolumn{2}{ |c| }{With NCR} \\ \hline
                 & Tree level & $\sigma_{stat}$ & EW+QED & $\Delta$ & EW+QED & $\Delta$ \\ \hline
 $\bar{\nu}_\mu$ & 12935 & 114 & 13850 &  915  & 13420 & 485 \\ \hline
 $\nu_\mu$       & 19947 & 141 & 18715 & -1232 & 19318 & -629 \\ \hline
    \end{tabulary}  
  \label{tab:totalEvents:Numu:AntiNumu:BESTrange}
\end{table}

The results for the neutrino beam mode are shown in
Table~\ref{tab:totalEvents:Numu:0.2}, where we see that $|\Delta|$ (with NCR) is smaller than for the antineutrino mode, Table~\ref{tab:totalEvents:AntiNumu:0.2}. Notice that the neutrino mode is expected to generate a larger number of events than the antineutrino one. However, $|\Delta|$ (with NCR) is smaller for neutrinos due to the radiative corrections sign's change. See Fig.~\ref{fig:corrections:nuAntinu}.

The other case of interest to consider is that of a threshold of
$0.7$~GeV, which corresponds to the already mentioned crossing point
for the neutrino mode. Starting from this energy, the radiative
corrections for the $\nu_\mu e^-$ are always negative, making the
effect for this case more visible than for the $0.2$~GeV
threshold. See
Table~\ref{tab:totalEvents:Numu:AntiNumu:BESTrange}. Since the
neutrino production is in general higher in this type of beams, we can
expect this result to be a general characteristic of this type of
experiments. Moreover, this shows the importance of setting up the
energy threshold based on the kind of physics measurement to be
conducted, instead of simply lowering the threshold based only in
detector characteristics.

\begin{figure}
\centering
   \subfloat[On-axis\label{fig:compa:0dega}]{% 
     \includegraphics[width=0.45\textwidth]{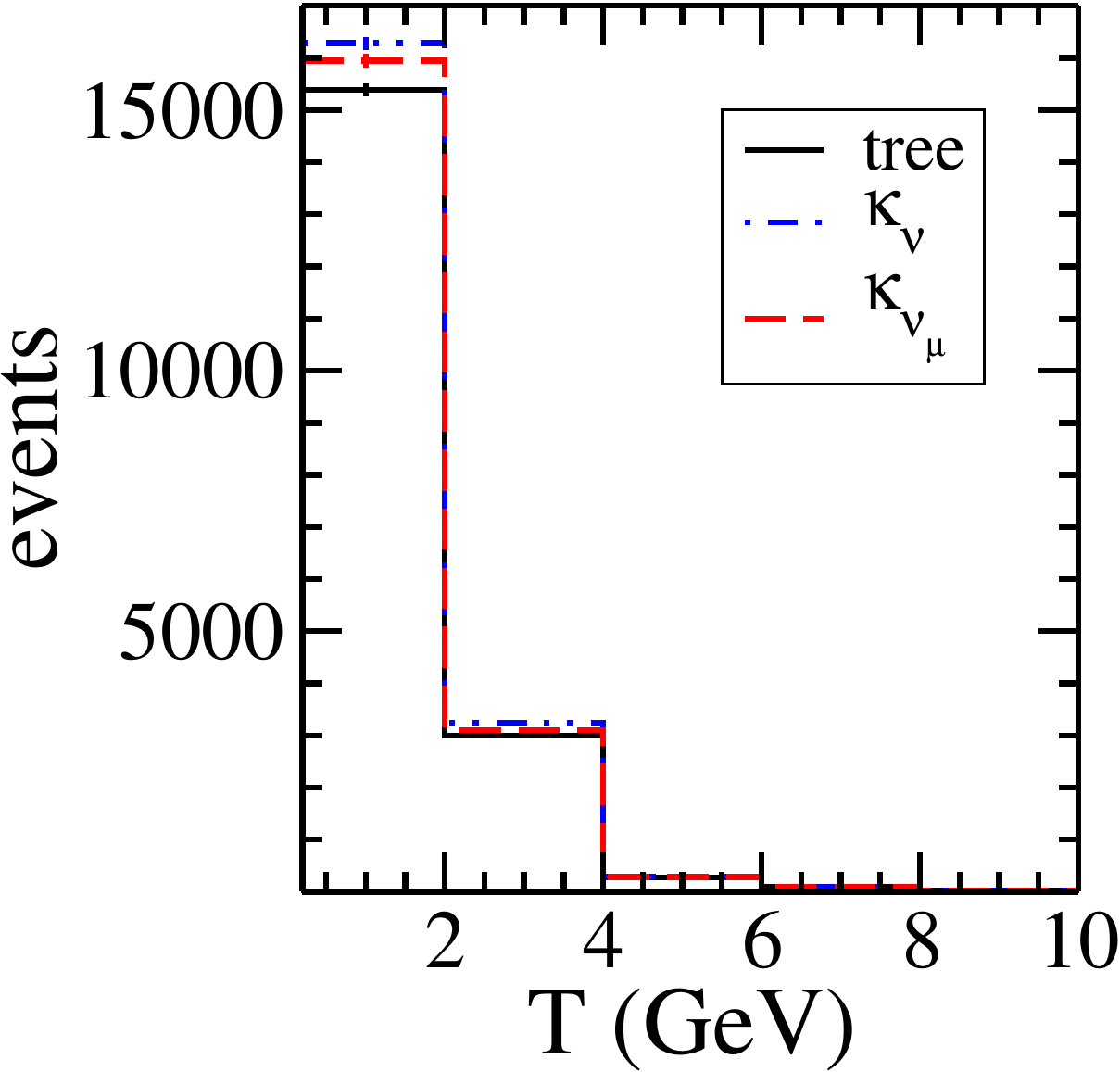}}
%  \hfill
   \subfloat[0.6$\degree$ off-axis \label{fig:compa:0.6dega}]{%
    \includegraphics[width=0.45\textwidth]{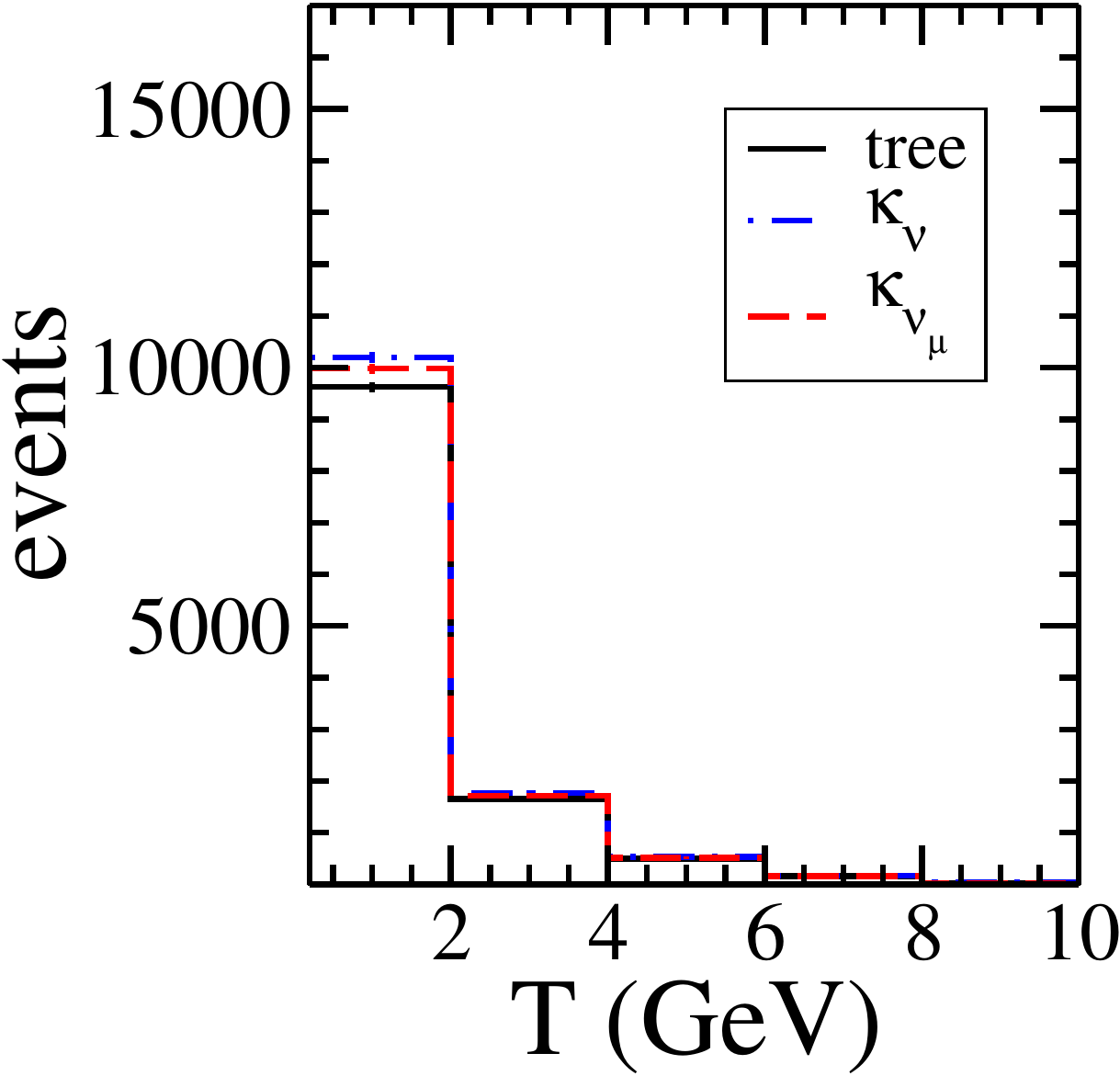}}
  \caption[]{\small Comparison among the number of $\bar{\nu}_\mu$ event expectations at tree-level
    (solid black line) and considering radiative
    corrections, with and without neutrino charge radius (dashed red
    and dot-dashed blue line, respectively). We show
    two DUNE-PRISM spectra: (a) On axis on the left and
    (b) 0.6$\degree$ on the right side.}
  \label{fig:dis:totalCorrections:AntiNumu}
\end{figure}

In Fig.~\ref{fig:dis:totalCorrections:AntiNumu}, we show the expected
number of events per bin of electron recoil energy for two different
incident antineutrino beam angles. The 2~GeV range for each bin is chosen conservatively, as the expected energy resolution for a DUNE ND-like detector is of the order of 10\% for energies above 0.2~GeV~\cite{AbedAbud:2021hpb}.
As already stated, there is 
better statistics when we consider the on axis position, which 
translates in a smaller error.  For antineutrino fluxes at other
angles, the statistics is worse, as is shown in the right panel of the
same Fig.~\ref{fig:dis:totalCorrections:AntiNumu}, for an angle of
$0.6\degree$. For larger angles, the statistics are even lower.  
We can also notice from this figure that the first energy bin shows the most relevant difference between the number of events for tree level and radiative correction expected measurements.
Finally, it is also evident from this figure that in the
antineutrino mode, a low energy threshold is very useful to have this
kind of signature.

\begin{figure}
\centering
   \subfloat[On-axis\label{fig:compa:0deg}]{% 
     \includegraphics[width=0.45\textwidth]{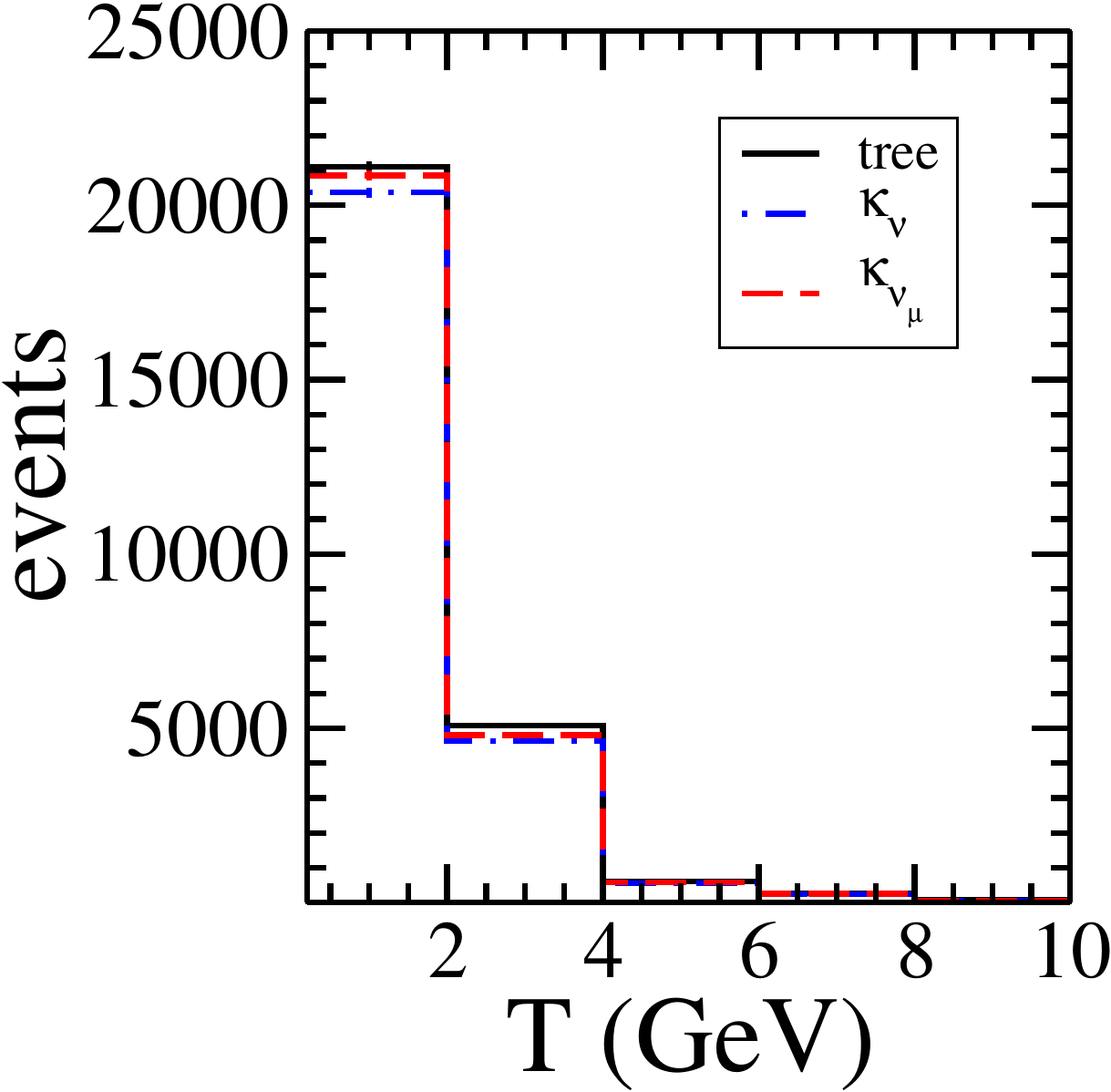}}
   \subfloat[0.6$\degree$ off-axis \label{fig:compa:0.6deg}]{%
    \includegraphics[width=0.45\textwidth]{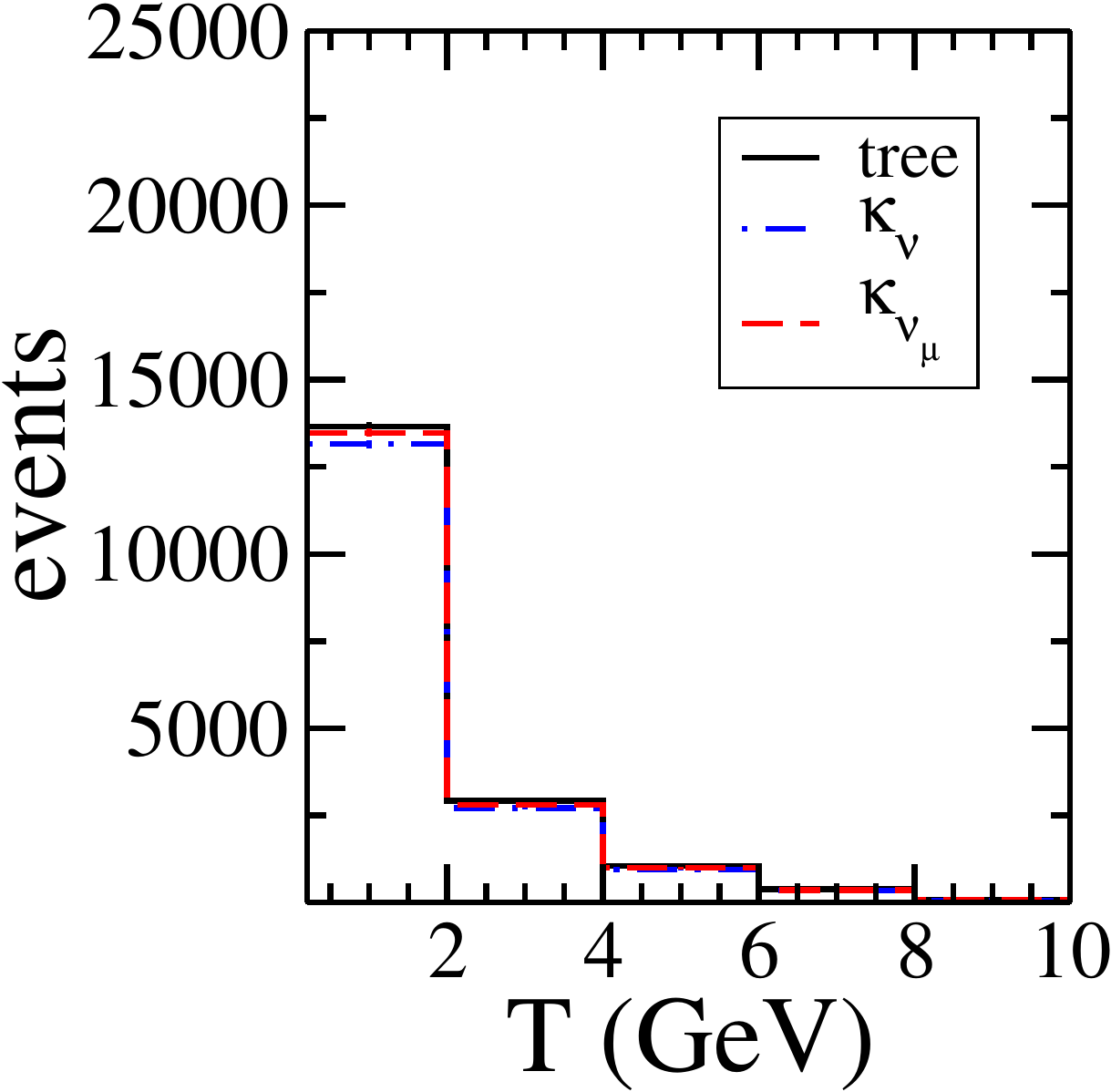}}
   \caption[]{\small Comparison among the number of $\nu_\mu$ event expectations at tree-level
    (solid black line) and considering radiative
    corrections, with and without neutrino charge radius (dashed red
    and dot-dashed blue line, respectively). We show
    two DUNE-PRISM spectra: (a) On axis on the left and
    (b) 0.6$\degree$ on the right side.}
    \label{fig:dis:totalCorrections:Numu}
\end{figure}

In Fig.~\ref{fig:dis:totalCorrections:Numu}, we show the results for
the case of muon neutrinos.  We notice that the radiative corrections
have an opposite contribution to the expected number of events as already
forecast. 
It is also
important to point out that the overall radiative correction
contribution is smaller for the neutrino mode than for the
antineutrino one. As already discussed, the reason is the cancellation
that occurs when considering two energy windows: the radiative
corrections in the neutrino mode change sign in the first bin of
energy; i.e., they have a positive contribution for energies below
approximately 0.7~GeV and a negative contribution for energies above
that limit.

\begin{figure}
\centering
   \subfloat[3\% systematics\label{fig:3-sigma}]{% 
     \includegraphics[width=0.5\textwidth]{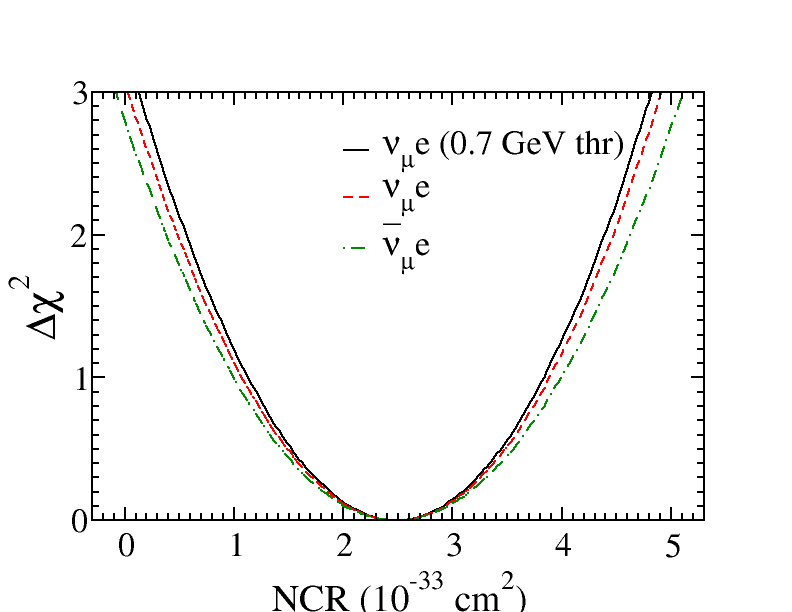} }
   \subfloat[5\% systematics\label{fig:5-sigma}]{% 
     \includegraphics[width=0.5\textwidth]{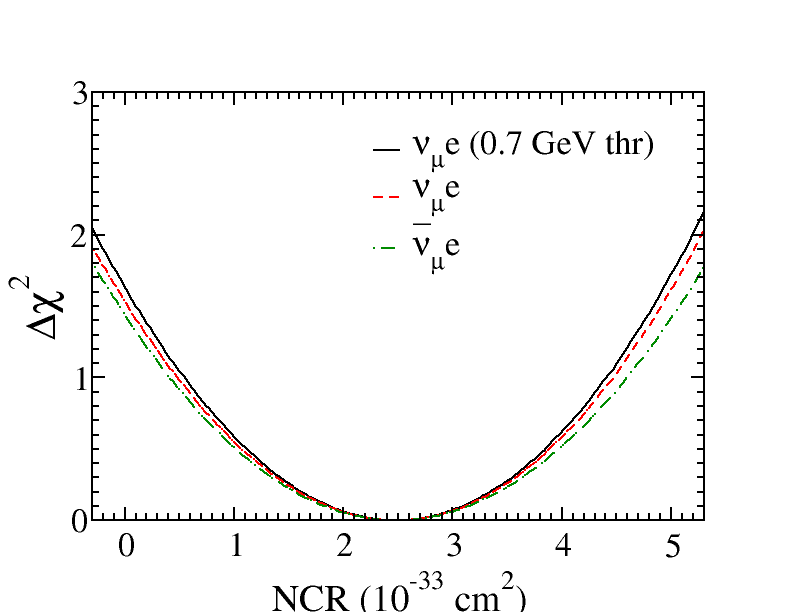} }  
   \caption[]{\small Expected sensitivity to the electroweak radiative corrections for a 3\% systematic error (a) Left; and 5\% systematic error (b) Right. We show $\Delta\chi^2=\chi^2-\chi^2_{\rm min}$ as a function of the neutrino charge radius (NRC). Red dashed and green dot-dashed lines correspond to a 0.2~GeV threshold in the $\nu$ and $\bar\nu$ scattering, while the black line corresponds to the 0.7~GeV $\nu$ scattering threshold. Neutrino data with 0.7~GeV threshold and 3\% systematic error can reach a better than 90\% confidence level sensitivity to the NCR within our assumptions. See text for details.}
\label{fig:chi1}
\end{figure}

Finally, to estimate the sensitivity to the radiative
  corrections, we conduct a $\chi^2$ analysis considering the expected
  number of events at a DUNE-PRISM like experiment with its
  statistical and systematic uncertainties. For this purpose, we assume
that the experiment will measure the SM prediction including radiative
corrections. We define the $\chi^2$ function as
\begin{equation}
\chi^2 = \sum_{i=1}^5 \dfrac{(N^{exp}_i-N^{theo}_i)^2}{(\sigma^2_{stat}+\sigma^2_{syst})_i} \,,
\end{equation}
  where $i$ is the energy bin, $N^{exp}$ refers to the expected number
  of events that the SM predicts, considering electroweak and QED
  radiative corrections, and $N^{theo}$ refers to the theoretically
  calculated number of events for different values of $\kappa$
  (Eq.~\ref{kappa:sirlin}).  The statistical and systematic
  uncertainties are given by $\sigma_{stat}$ and $\sigma_{syst}$,
  respectively.  We assume the statistical uncertainty to be the
  square root of the number of events, $\sigma_{stat}=\sqrt{N^{exp}}$,
  and the systematic to be 3\% or 5\% error. We also define
  $\Delta\chi^2=\chi^2-\chi^2_{\rm min}$, where $\chi^2_{\rm min}$ is
  the minimum value of $\chi^2$. 
  
  The results with different systematic uncertainties are depicted in Fig.~\ref{fig:chi1},
  where we see that it may be possible to distinguish the
  prediction $\kappa_{\nu_\mu} = 0.9925$, for radiative corrections with NCR, from the case without NCR, $\kappa_\nu
  = 1.0176$.
    A $3\%$ systematic error would be sensitive, at
    $1\sigma$ precision, to a NCR in the range from $1.0$ to
    $4.0$~$\times 10^{-33}$~cm$^2$ for the antineutrino channel, and
    from $1.1$ to $3.9$~$\times 10^{-33}$~cm$^2$ for the neutrino
    channel, and, finally, from $1.1$ to $3.8$~$\times
    10^{-33}$~cm$^2$ for the neutrino channel with the energy
    threshold of $0.7$~GeV.  Even in the case of a $5$\% systematic
    error, it is still possible to have a precision higher than
    $1\sigma$. In contrast, the current constraint reported in the PDG
    for $\nu_\mu e$ scattering is in the range from $-5.3$ to
    $6.8$~$\times 10^{-33}$~cm$^2$~\cite{Zyla:2020zbs}, which is still
    consistent with no NCR.

\begin{figure}
\centering
    \includegraphics[width=0.45\textwidth]{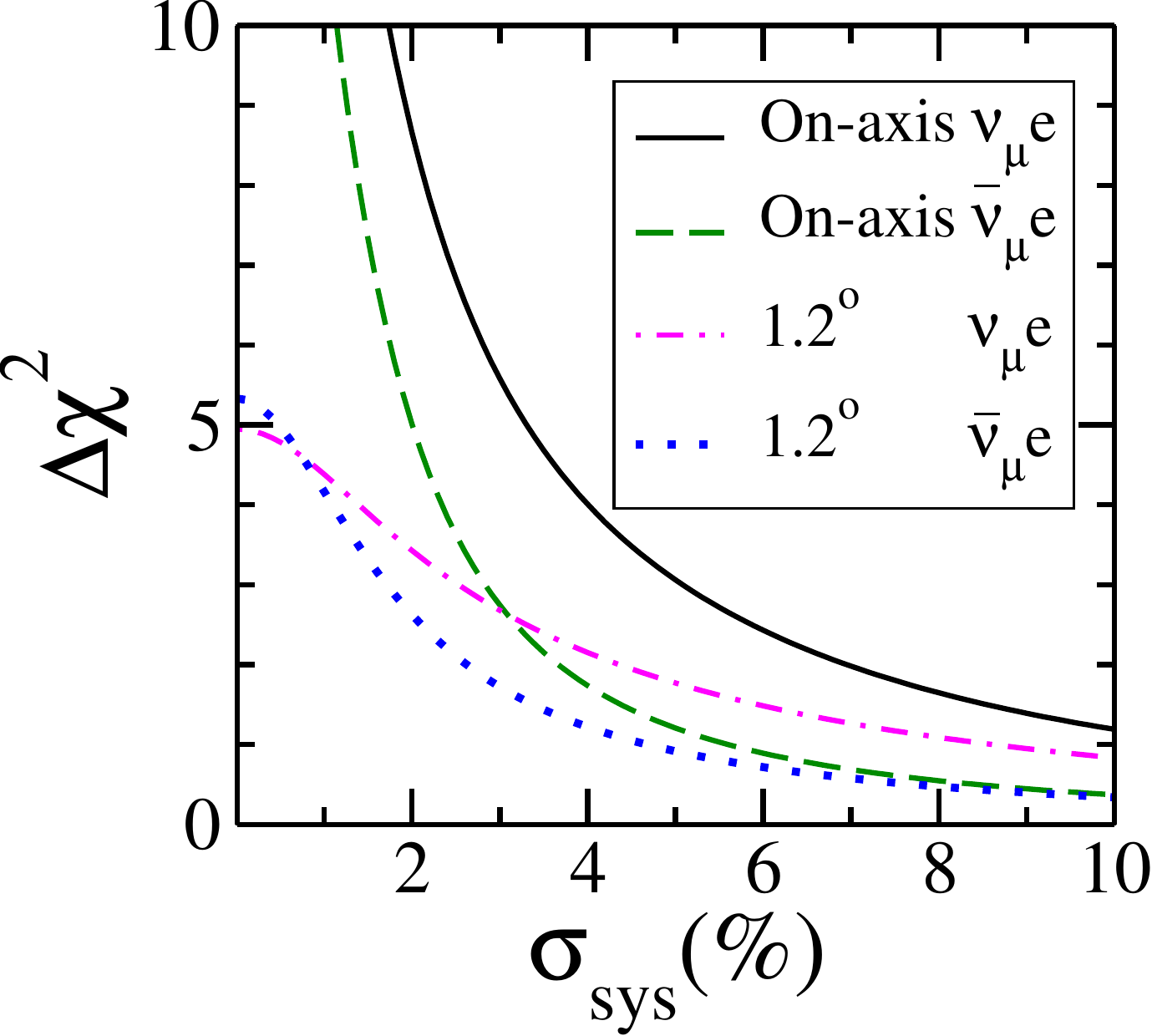}%}
    \caption[]{\small Expected sensitivity, in terms of
      $\Delta\chi^2=\chi^2-\chi^2_{\rm min}$, to differentiate between
      the tree level and the radiative corrections case, depending on
      the systematic error.  We show the results for two different
      locations of the detector (on axis and $1.2^\circ$) and for neutrino
      (solid and dash-dotted) and antineutrino (dashed and dotted)
      electron scattering. For the neutrino electron case, we have
      chosen an energy threshold of $0.7$~GeV to improve the
      sensitivity.  }
\label{fig:chi2}
\end{figure}

    Besides this analysis, we have also performed a different
    computation shown in Fig.~(\ref{fig:chi2}). For this computation we
    consider the $\Delta\chi^2$ with a theoretical
    prediction where no radiative corrections are taken into account
    at all, that is, the tree level. We show in this figure that,
    depending on the systematic uncertainties, the radiative
    corrections can be distinguished from the tree level or not. For
    the neutrino case we consider, as discussed above, an appropriate
    energy range from $0.7$ to $10$~GeV to improve the sensitivity,
    while for the antineutrino case we consider an energy range
    starting from $0.2$~GeV. We have considered five energy bins in
    both cases. The results are shown for two different incoming
    neutrino angles. We can notice that the neutrino case is very
    promising in its sensitivity to radiative corrections even for
    relatively large systematic errors and for different incoming
    neutrino fluxes thanks to the combination of high statistics and a
    suitable energy window. For the antineutrino case it is also
    possible to have a discrimination, but the systematic errors
    should be under control, approximately below 4\%.

    We could have even better discrimination of the radiative
    corrections if we combine both neutrino and antineutrino
    signals. We notice that radiative corrections
      have opposite effects on neutrino and antineutrino electron
      scattering, resulting in a cross section decrease for muon
      neutrino scattering off electrons but an increase for
      antineutrino interactions.
      This suggests that it might be
      possible to define the difference between these two signals as
      an observable to evaluate the neutrino charge radius better.
      A detailed analysis
    in this direction would require a good knowledge of the
    correlation between both signals.

\section{Conclusions}

The precise
determination of the radiative corrections at low energies is of great
importance to test the SM. An accurate determination of the
weak mixing angle in the low energy region of accelerator-based neutrino experiments is in order, as well as an
experimental probe that the neutrino charge radius (NCR), as an effective observable, is
present in this process. Moreover, accurate tests of physics beyond the
SM will find a limitation if these observables are not well
determined.

We have studied the sensitivity of future near detectors, like in
long-baseline neutrino experimental facilities, to radiative
corrections in neutrino-electron scattering, considering the case of
DUNE-PRISM as an illustrative example. We focus on the NCR as an
effective observable that is characteristic of this process. Since the
NCR main effect is a shift in the weak mixing angle, we have
investigated the detector sensitivity to radiative corrections,
separating the NCR effect.  Taking as a guidance, the DUNE-PRISM
configuration that would allow several beam angle setups, we have
analyzed different neutrino energy spectra. We find that on axis
neutrino spectrum will allow a better determination of the radiative
corrections and possibly the NCR due to its higher statistics. We have
illustrated that with a systematic error of the order of $3$\% there are
good expectations to measure the NCR with an error of the order of $1.5\times10^{-33}$~cm$^2$. 

Our analysis shows that for the case of a $\nu_\mu$ beam, a correct
selection of the energy window could allow us to determine the
existence of the NCR if the systematic uncertainties are under
control, thanks to the high statistics expected in this beam mode.
On the other hand, for the $\bar{\nu}_\mu$ mode, we have pointed out
that the best chance to measure this effective observable is for small
electron recoil energy values. Therefore, in this case, the lower the
threshold, the better for such a measurement.

\section*{Acknowledgments}
This work was supported by CONACYT-Mexico Grant No. A1-S-23238, SNI (Sistema Nacional de Investigadores). CAM acknowledges support from FAPESP Grant Process No. 2014\-/\-1\-9\-1\-6\-4-6.

\appendix

\section{QED Functions}
\label{appen.A}

In this appendix, we show the explicit form of the functions
$f_{-}(z)$, $f_{+}(z)$, and $f_{+-}(z)$ that are introduced in
Eq.~(\ref{eq:cs:ewqed}). We consider the expressions given in
Ref.~\cite{Bahcall:1995mm} (numerical expressions can be found in
Ref.~\cite{Passera:2000ug}) and that for the case of $f_{-}(z)$ is

\begin{align}
  \label{eq:f-}
  f_{-}(z) 
  &= \left[ \frac{E}{l}\ln\left( \frac{E+l}{m_e}\right)-1 \right] \left[ 2\ln\left( 1 - z -\frac{m_e}{E+l} \right) -\ln\left( 1-z \right) -\frac{1}{2}\ln z - \frac{5}{12}\right] \nonumber\\
  &\qquad {} +  \frac{1}{2}\left[ L(z) - L(\beta) \right] - \frac{1}{2}\ln^2\left( 1-z \right) -\left( \frac{11}{12} + \frac{z}{2}\right) \ln \left( 1-z \right) \nonumber\\
  &\qquad {} +  z\left[\ln z + \frac{1}{2}\ln\left( \frac{2E_{\nu}}{m_e}\right)\right] - \left( \frac{31}{18} + \frac{1}{12}\ln z \right)\beta -\frac{11}{12}z + \frac{z^2}{24} \,,
\end{align}
where $l=\sqrt{E^2-m^2_e}\,$ is the three-momentum of the electron,
$E=T+m_e$, $\beta=l/E$, and $L(x)$ is in Spence's function space
corresponding to the following dilogarithm:

\begin{equation}
  L(x)=-Li_2(x)=\int_0^x \frac{\ln \abs{1-t}}{t}dt \,.
\end{equation}

The $f_{+}(z)$ function is given by

	\begin{align}
  	 \label{eq:f+}
	    \left( 1-z\right)^2 f_{+}(z) 
	    &= \left[ \frac{E}{l}\ln\left( \frac{E+l}{m_e}\right)-1 \right] \left\lbrace \left( 1-z\right)^2 \left[2\ln\left( 1 - z -\frac{m_e}{E+l} \right) \right. \right. \nonumber\\
	&\qquad {} \left. \left. -\ln\left( 1-z \right) -\frac{1}{2}\ln z - \frac{2}{3}\right] - \frac{z^2\ln z + 1-z}{2} \right\rbrace \nonumber\\
    &\qquad {} -  \frac{\left( 1-z\right)^2}{2}\left\lbrace \ln^2 \left( 1-z\right) + \beta \left[ L(1-z) - \ln z \ln \left( 1-z\right) \right] \right\rbrace  \nonumber\\
    &\qquad {} + \ln \left( 1-z\right) \left[ \frac{z^2}{2}\ln z + \frac{1-z}{3} \left( 2z-\frac{1}{2} \right)  \right] -\frac{z^2}{2} L\left(1-z \right)  \nonumber\\
    &\qquad {} - \frac{z \left( 1-2z\right) }{3} \ln z - \frac{z\left( 1-z\right)}{6} \nonumber\\
    &\qquad {} - \frac{\beta}{12}\left[\ln z + \left( 1-z\right)\left( \frac{115-109z}{6}\right) \right] \,,
    \end{align}
and the $f_{+-}(z)$ function is
	\begin{align}
  	 \label{eq:f+-}
	    f_{+-}(z) 
	    &= \left[ \frac{E}{l}\ln\left( \frac{E+l}{m_e}\right)-1 \right] 2\ln\left( 1 - z -\frac{m_e}{E+l} \right) \,.
    \end{align}

\cleardoublepage

\addcontentsline{toc}{chapter}{Bibliografía}
\markboth{BIBLIOGRAFÍA}{BIBLIOGRAFÍA}
\bibliographystyle{ieeetr} % estilo de la bibliografía. acm
\bibliography{bibliografia} % texto.bib es el fichero donde está salvada la

\end{document}